\documentclass[12pt]{article}
\begin{document}
\setcounter{page}{0}
\input epsf
\title{Retrieval and Chaos in Extremely Diluted Non-Monotonic Neural Networks}
\author{M. S. Mainieri and R. Erichsen Jr.\\
\small
Instituto de F{\'\i}sica, Universidade Federal do Rio Grande do Sul,\\
\small
Caixa Postal 15051, CEP 91501-970, Porto Alegre, RS, Brazil}
\maketitle

\thispagestyle{empty}
\begin{abstract}
We discuss, in this paper, the dynamical properties of extremely diluted,
non-monotonic neural networks. Assuming parallel updating and the Hebb
prescription for the synaptic connections, a flow equation for the macroscopic
overlap is derived. A rich dynamical phase diagram, was obtained, showing a
stable retrieval phase, as well as a cycle two and chaotic behavior. Numerical
simulations were performed, showing good agreement with
analytical results. Furthermore, the simulations give an additional insight
into the microscopic dynamical behavior during the chaotic phase. It is shown
that the freezing of individual neuron states is related to the structure
of chaotic attractors.
\end{abstract}



\newpage
\label{sec:level1}

\section*{1. Introduction}

\noindent
Models of neural networks have been largely studied since the pioneering work
of Hopfield \cite{Hopfield}. Assuming symmetric interactions and monotonic
units, and using classical tools of statistical mechanics, equilibrium
properties were extensively investigated. For a review in this subject, see,
e. g., \cite{Hertz}. Nevertheless, the brain is a highly dynamical system, and
equilibrium statistical mechanics should be not able to describe most of its
properties. At our knowledge, the first work including the dynamical
behavior as an important topic in neural networks research is due to
Sompolinsky et al. \cite{Sompolinsky}. In that work, it was shown that
asymmetric synaptic connections between units, a plausible feature, results in
cyclic and chaotic behavior in the network dynamics. The chaotic behavior in
neural tissue is, in fact supported by biological observations \cite{Nicolis}.

More recently, a new class of models allowing for the complex behavior in
neural networks
started to be explored. It consists in networks of neurons in which the
output activity is a non-monotonic function of the local input field. 
Next, we present a quick list with some important results available in
the literature. 

The feed-forward perceptron is an architecture where the main task to be
performed is the classification of a set of input patterns. Following first
step replica symmetry breaking calculations in the space of the synaptic
interactions, see \cite{BMZ,BE95} and references therein, the
optimal capacity for classification of patterns is larger for the
non-monotonic than for the monotonic perceptron. Still concerning this
subject, non-monotonic perceptrons offer a further opportunity to study the
fractal organization of the space of interactions \cite{MO}.

In networks of recurrent architecture, the main subject concerns the
associative memory property. If the dynamical evolution is governed by an
hamiltonian, with the relaxation to an equilibrium distribution of states
reproducing the process of memory retrieval. The
question here is, if the dynamics is not governed by an hamiltonian, as in the
case of non-monotonic, under which conditions it evolves to an equilibrium
distribution of states keeping, in this way, the associative memory property? 
This problem was studied by Inoue \cite{JI96}, through the use of an
equilibrium signal to noise analysis. It was found that, depending on the
parameters defining the non-monotonicity, for binary patterns, and using the
Hebb learning rule, the storage capacity, to be defined and discussed later,
is highly improved, compared to monotonic neurons.

The dynamics of extremely diluted networks of continuous, non-monotonic
neurons was studied in \cite{BV93}. The associative memory property is still
present, and a non-regular, chaotic-type attractor was also noticed. The
authors pointed out the emergence of chaos as a consequence of the
non-monotonicity. The 
categorization ability of extremely diluted, three-state, non-monotonic
neural network was studied in \cite{DT96,DD96}. A categorization phase, as
well as a chaotic phase, was noticed. In \cite{CMNS99} there is a study of
the chaotic attractors of a model with non-monotonic, binary neurons, with
finite connectivity and binary synapses.

Our purpose in this paper is to study some of the dynamical properties of
non-monotonic neural networks.  The relaxation to the retrieval
state, the appearing of oscillations and chaos, and the interplay between all
these behaviors are discussed. In order to make transparent
the essential physical aspects, the simplest model of non-monotonic neural
network was chosen, i. e., the network of binary, reverse-wedge
neurons, with hebbian connections. In order to have an analytically solvable
dynamics, we
restrict our study to the extremely diluted architecture with parallel
updates. In addition, further insight on the microscopic dynamics
during the chaotic phase, as well as in the transition from the chaotic
to the retrieval phase, was obtained through numerical simulations.

\section*{2. Diluted network dynamics}

We consider a network composed by N sites, whose state is represented by a set
of variables $\{S_i\}\;i=1\dots N$, with the state of the neuron in site $i$
being represented by a binary variable $S_i=\pm 1$. In the absence of retrieval
noise, each neuron updates in a synchronous, deterministic way according to
\begin{equation}
S_i(t+1)=F_{\theta}\left(h_i(t)\right)\;,
\label{1}
\end{equation}
where
\begin{equation}
F_{\theta}(h)=\left\{
	\begin{array}{
	r@{\quad:\quad}l} +1 & h<-\theta\;\; 
	{\rm and}\;\; 0<h<\theta\\ -1 & {\rm otherwise}
	\end{array} \right.
\label{2}
\end{equation}
is the reverse-wedge activation function. Here, $\theta$ is a threshold
parameter. The local field felt by the neuron at site $i$ is given by
\begin{equation}
h_i(t)=\sum_{i=1}^N J_{ij}S_j\;.
\label{3}
\end{equation}
In the limit of extreme dilution each neuron is connected to $C$ other
neurons, with 
\begin{equation}
	1\ll C\ll \ln N\,.
\label{3.1}
\end{equation}
Due to this condition, the synaptic connections have a
tree-like structure, and the existence of feedback-loops has a vanishing
probability \cite{BV93}. In consequence, the dynamics of this network is
exactly solvable \cite{DGZ}. The storage capacity of the network is defined as
$\alpha=p/C$. We assume that the synapses are given by a Hebb-like learning
rule,
\begin{equation}
J_{ij}=\frac{C_{ij}}{C}\sum_{\mu=1}^p \xi_i^{\mu}\xi_j^{\mu}\;,
\label{4}
\end{equation}
where $C_{ij}$ are random binary variables assuming values $1$, if
the output of neuron $j$ is connected to the input of neuron $i$ and zero,
otherwise. They are chosen according the probability distribution 
\begin{equation}
P(C_{ij})=\frac{C}{N}\delta(C_{ij}-1)
	+\left(1-\frac{C}{N}\right)\delta(C_{ij})\;,
\label{5}
\end{equation}
where $\delta(x)$ is the Dirac's delta function. The memories
$\left\{\xi_i^{\mu}\right\}\;,\;\;i=1\dots N\;,\;\;\mu=1\dots p$ are a set of
independent, identically distributed random binary variables, chosen according
the probability distribution 
\begin{equation}
P\left(\xi_i^{\mu}\right)=\frac{1}{2}\delta\left(\xi_i^{\mu}-1\right)
	+\frac{1}{2}\delta\left(\xi_i^{\mu}+1\right)\;.
\label{6}
\end{equation}
The only relevant order parameters describing the macroscopic state of the
network is the set of overlaps
\begin{equation}
m_{\mu}=\frac{1}{N}\sum_{i=1}^N \xi_i^{\mu}S_i
\label{7}
\end{equation}
between the system configuration the stored patterns.

In what follows, we consider that only one memory pattern, say $\mu=1$,
has a macroscopic overlap of order ${\cal O}(1)$, and all the others $p-1$ are
of order ${\cal O}(1/\sqrt{N})$. Following the now standard derivation of
\cite{DGZ}, we obtain that, in the absence of retrieval noise, the retrieval
overlap satisfies the recursion relation
\begin{equation}
m_1(t+1) = - \frac{1}{2} {\rm erf} \left(\frac{m_1 (t) + \theta}
	{\sqrt{2\alpha}}\right) + \frac{1}{2} {\rm erf} \left(\frac{m_1 (t)}
	{\sqrt{2\alpha}}\right) - \frac{1}{2} {\rm erf} \left(\frac{m_1 (t)
	- \theta}{\sqrt{2\alpha}}\right)
\,.
\label{8}
\end{equation}
For parallel dynamics, in the limit of extreme dilution, this equation
fully describes the dynamical evolution of the network.

Before to extend the discussion for all $\theta$, it worths to mention what is
the behavior of the system in the limits $\theta\rightarrow\infty$ and
$\theta\rightarrow 0$. For $\theta\rightarrow\infty$, the deterministic neuron
update reduces to the 
signal transfer function $S_i (t + 1) = {\rm sgn} [h_i (t)]$. The macroscopic
dynamical evolution is given by $m (t + 1) = {\rm erf} (m (t) / \sqrt{2
\alpha})$. According to \cite{DGZ}, this equation has a fixed-point for
$t\rightarrow\infty$ at a finite value $m = m (\alpha)$, provided that $\alpha
< 2/\pi$. At this value of the storage capacity, the system undergoes a
continuous transition to the fixed point $m = 0$. 

In the limit
$\theta\rightarrow 0$, the neurons update according to the transfer function
$S_i (t + 1) = - {\rm sgn} [h_i (t)]$, with the macroscopic dynamics being
given by $m (t + 1) = - {\rm erf} (m (t) / \sqrt{2 \alpha})$. This equation no
longer have a finite fixed-point for finite $m$. Now, for small
$\alpha$, the network evolves dynamically through a cycle of length two, where
the finite retrieval overlap changes from $m$ to $- m$ from one time step to
the other. In other words, in 
one time step, it retrieves one pattern, and in the next time step it
retrieves the inverse of this pattern. The amplitude
$m$ of the limit cycle decreases continuously with the increasing $\alpha$,
arriving to $m = 0$ at the same storage capacity $\alpha = 2/\pi$ as in the
previous case $\theta\rightarrow\infty$.

Now we start to discuss the long term behavior of the system. 
A complete dynamical phase diagram for general $\theta$ is shown in
figure~1. In 
region I, corresponding to large $\alpha$, the only fixed point of
Eq.~(\ref{8}) is $m = 0$, known as the zero solution \cite{BSVZ94}. The
region II, for intermediate and large $\theta$, is the locus of the retrieval
solution, with finite fixed point overlap $m\neq 0$. The
transition between regions I and II is continuous. It worths to remark that
the borderline for the retrieval region decreases monotonically from
$\alpha=2/\pi$ for large $\theta$ to $\alpha=0$ for
$\theta=0$. This is in contrast to that reported by Inoue \cite{JI96} for
the fully connected network. There, the Hopfield limit $\alpha\approx 0.138$
is found for large $\theta$. Then, decreasing $\theta$, the borderline for the
retrieval increases up to a maximum $\alpha\approx 0.211$ at $\theta\approx
1.77$. For smaller $\theta$, it decreases till to reach $\alpha=0$ at
$\theta=0$.

The region III is divided in two parts, one for small $\theta$ and
$\alpha<2/\pi$ and other close to the the $\theta$ axis, for $\theta<1$. In
this region, the network shows the cycle two behavior, as in the
$\theta\rightarrow 0$ limit. The amplitude of the cycle two decreases
monotonically to zero at the transition line between regions I and III. The
transition from region III to other region than I is discontinuous.

In region IV the network has a periodic behavior. Leaving the retrieval
region, roughly at the right and above, in the figure, it starts a
period doubling cascade that ends in the chaotic region V.

Finally, we discuss the regions that are signaled with a star in figure~1. The
star means that the long term behavior depends on the initial value $m_0$ of
the overlap. If the initial value is sufficiently small, the network evolves to
the basin of attraction of the corresponding phase without the
star. Otherwise, it ends in the cycle two behavior, like in phase III.

\section*{3. Numerical simulations}

The non-monotonic network with extremely diluted connections offers an
interesting opportunity to compare microscopic and macroscopic views of
non-linear phenomena. As an example, the bifurcation diagram of the overlap $m$
for $\alpha=0.04$ and varying $\theta$ is shown in figures 2a and 2b,
determined, respectively from the macroscopic map of Eq.~(\ref{8}) and the
microscopic definition of Eq.~(\ref{7}). The initial condition
is $m_0=0.1$, i.~e., outside the basin of attraction of the cycle two
solution. Figure 2b results from
simulation over a network with $N=10000$ and $C=100$. In order to save
computer memory, only nonzero elements of the
synaptic matrix where stored, as in \cite{AL}. Comparing the two diagrams, we
can see that bifurcations and periodicity windows are less clearly defined in
the simulation than in the macroscopic evolution. Furthermore, in the range in
$\theta$ where  
states with $m>0$ and $m<0$ coexist in the same attractor, the diagram from
simulation is more populated with states with small overlap. This is in
contrast with the macroscopic evolution, that results in a more uniformly
distributed diagram. We may ask if these discrepancies are due to the high
connectivity value, since $C/N=0.01$ in figure 2b, while figure 2a holds in the
limit $C/N\rightarrow 0$. We believe that this is not the case, since
simulations with $N=1000$ and $C=100$ give similar results. Instead of this,
we believe that the discrepancies are due to finite size effects, since the
definition $\alpha\equiv p/C$ holds for both $p$ and $C$ going to $\infty$
while, in the simulations we have $p=4$. Nevertheless,
although the  discrepancies, there is a good general agreement between the two
diagrams. In particular, the figures agree in which concerns to transitions
from cyclic to chaos and to retrieval regions. This means that, even if
Eq.~(\ref{3.1}) is not strictly satisfied, the simulations reproduce quite
well the dynamics of the diluted network. With this, we conclude that it is
justified to proceed with simulations, in order to obtain additional insight
about the chaotic behavior of 
the network. For completeness it is shown, in figure 2c, the Lyapunov
exponent calculated from iterations of the macroscopic map (Eq~\ref{8}). The
positive values assumed confirm that the system displays a chaotic phase.

The description of the dynamical evolution through macroscopic variables
leaves open some questions, mainly concerning the chaotic regime. We could ask,
for instance, if the abrupt change in the overlap from step to step results
from alternate flipping activity in all network sites, or if part of the
sites are frozen and part are flipping. In order to have some insight about
this and other questions, computational simulations were realized. Here we
introduce, for
each neuron $i$, the quantity $w_i (t)\equiv t - t_i$, where $t_i$ is the
time corresponding to the last flip of neuron $i$. Expressed in words,
$w_i(t)$ is the number of time steps that neuron $i$ stays in the same state,
at time step $t$. Suppose that the
network is in the cycle of period 2 with maximal amplitude. Then, all sites
flip in each time step. In this case, $w_i = 0$ for all site $i$. Now
suppose that, after to start, the network quickly arrives to an equilibrium
state with all the neurons frozen in one state. In
this case, after the time step $t$, all $w_i$ are smaller but close to
$t$. Between these two extreme, the distribution $P(w)$ gives a good picture
of the dynamical state of the network. 

The results for $P(w)$ are shown in figures~3a and 3b. To understand the
figures, it is sufficient to know that if a neuron had its last flip $n$ time
steps before $t$, then it contributes with a count for $w=n$. The final time
we adopted is $t=500$, since this is a sufficient amount of time to reach the
long-term behavior. Each figure results from an average over $50$
runs. To take care of finite size effects, simulations were done for $N=5000$
and $N=10000$. Since the results were equivalent, they are shown only
for $N=10000$. We also have $C=100$ and $p=4$, so $\alpha=0.04$. The
initial condition is $m_0=0.1$. Different regions of the phase diagram were
visited through changing $\theta$. In figure~3a, $P(w)$ is shown for
$\theta<1.0$. The main feature to be extracted is that no freezing occurs:
all neurons flip in a time interval smaller than $50$ time-steps. For
$\theta=0.25$, the network in the region ${\rm I}^*$ of the phase diagram,
inside the basin of attraction of the zero phase. The condition $\theta=0.3$
corresponds to a network in region ${\rm II}^*$, inside the basin of attraction
of the retrieval phase, with a retrieval overlap $m \approx 0.1$ (see
figure~2). Since the values of overlap involved are small, we could expect
that part of neurons remains frozen, but this is in contrast to what actually
happens. When $\theta=0.7$, the network shows chaotic behavior. Since the
chaotic attractor wanders through states with $m$ going from large negative to
large positive values (see figure~2b),
we could expect, as it is confirmed, that all neurons have flipping
activity. For these three values of $\theta$, $P(w)$ shows an
exponential decaying behavior, with a similar decaying rate. Results for
$\theta=0.9$ is also included in figure~3a. Now we observe a distinct
behavior, with $P(w)$ decaying more slowly than in the previous cases. We
guess that this is already a consequence of the dynamical transition that
occurs at a slightly higher $\theta$ value, where the chaotic attractor splits
in two, one for $m>0$ and other for $m<0$, as can be seen in figure~2b. 

The plot of $P(w)$ for $\theta=1.0$, $\theta=1.2$ and $\theta=1.3$ is
shown in figure~3b. For $\theta=1.0$, the system is in phase V, just above
the $\theta$ value where the chaotic attractor splits in two.
The decaying is slower than exponential, but there are not
frozen sites. We conclude that the splitting in the chaotic attractor is not
related to the freezing of a part of sites. Rather, it is related to a
change in the decaying rate in $P(w)$ that allows for an important decrease in
the flipping activity. When $\theta=1.2$, the network
is in the periodic regime of phase IV, with positive overlap. The flipping
activity decreases, as reveals the small population for
$w\approx 0$, and an important concentration of frozen sites can be seen for
large $w$. In view of this, we may state that the appearing of a periodic
retrieval regime with positive $m$ is related to the freezing of part of the
neurons. It worths to note that $P(w)$ shows a nontrivial structure for
$w>300$. We don't know, at this moment, how to explain this structure. Finally,
for $\theta=1.3$ the system is in retrieval regime. The figure shows a small
population for small $w$ corresponding to sites that are flipping. There is a
range for intermediary $w$ with zero or very small population, and a large
population for large $w$, corresponding to frozen sites. Since the retrieval
overlap $m\approx 0.93$ is large, we conclude that most of sites are frozen in
the direction of the pattern.

\section*{4. Conclusion}

Following a dynamical approach, we have studied the extremely diluted
non-monotonic neural network. The results show that  there is no improvement
in the storage capacity of the retrieval phase due to the non-monotonicity: it
is always smaller than that of the network with monotonic neurons. As stated
above, this is in contrast with the highly connected network
\cite{JI96}. There, an optimal $\theta$, whose storage capacity is larger than
that of  the monotonic network, was observed. It would be interesting to study
a network with finite connectivity, ranging from zero, that is the present
model, to 1, that is the highly connected model. In this way, it would be
possible to localize where is the changeover from one regime to the other.

Numerical simulations were performed in order to get additional information
about the microscopic dynamical evolution. The quantity observed was
the amount of time that each neuron $i$ is fixed without flip,
$w_i$. Excluding the trivial cycle-two regime for $\theta<0.2$, where
$P(w)\approx\delta(w)$, the distribution $P(w)$ shows three distinct
regimes, depending on the flipping activity of the network: high flipping
activity with exponential decaying $P(w)$, mean flipping activity with
less than exponential decay, and low flipping activity with freezing.

We conclude that the distribution $P(w)$ gives a good
picture of the microscopic evolution of the network, in its different dynamical
regimes. It was not the aim of this paper, but in a future work it could be
interesting to investigate quantitatively the decaying law in the
second regime.

\section*{Acknowledgements}

The authors thank Dr. W. K. Theumann, Dr. M. A. P. Idiart, Dr. F. A Tamarit and
Dr. N. Lemke for fruitful discussions. M. S. M. thanks CNPq (Conselho
Nacional de Desenvolvimento Científico e Tecnológico - Brazil) for financial
support.

\newpage

\section*{Figure Captions}

\noindent
{\bf Figure 1:} Phase diagram for the extremely diluted non-monotonic neural
network. Solid (dotted) lines correspond to discontinuous (continuous) phase
transitions. For the dynamical behavior in each phase, see the text.\\

\noindent
{\bf Figure 2:} Bifurcation diagram and Lyapunov exponent for the extremely
diluted non-monotonic neural network, for $\alpha=0.04$ and varying $\theta$,
with the initial condition $m_0=0.1$. (a) Bifurcation diagram obtained from
iteration of Eq.~(\ref{8}); (b) bifurcation diagram obtained from simulation
in a network with $10000$ neurons, computing the overlap in each time step
from its definition in Eq.~(\ref{7}); (c) Lyapunov exponent, obtained from the
iteration of Eq.~(\ref{8}).\\

\noindent
{\bf Figure 3:} Distribution $P(w)$ for the extremely diluted non-monotonic
neural network, for $\alpha=0.04$, after simulations with $10000$ neurons. (a)
threshold $\theta$ in the range $0.25$ -- $0.9$; (b) threshold $\theta$ in the
range $1.0$ -- $1.3$

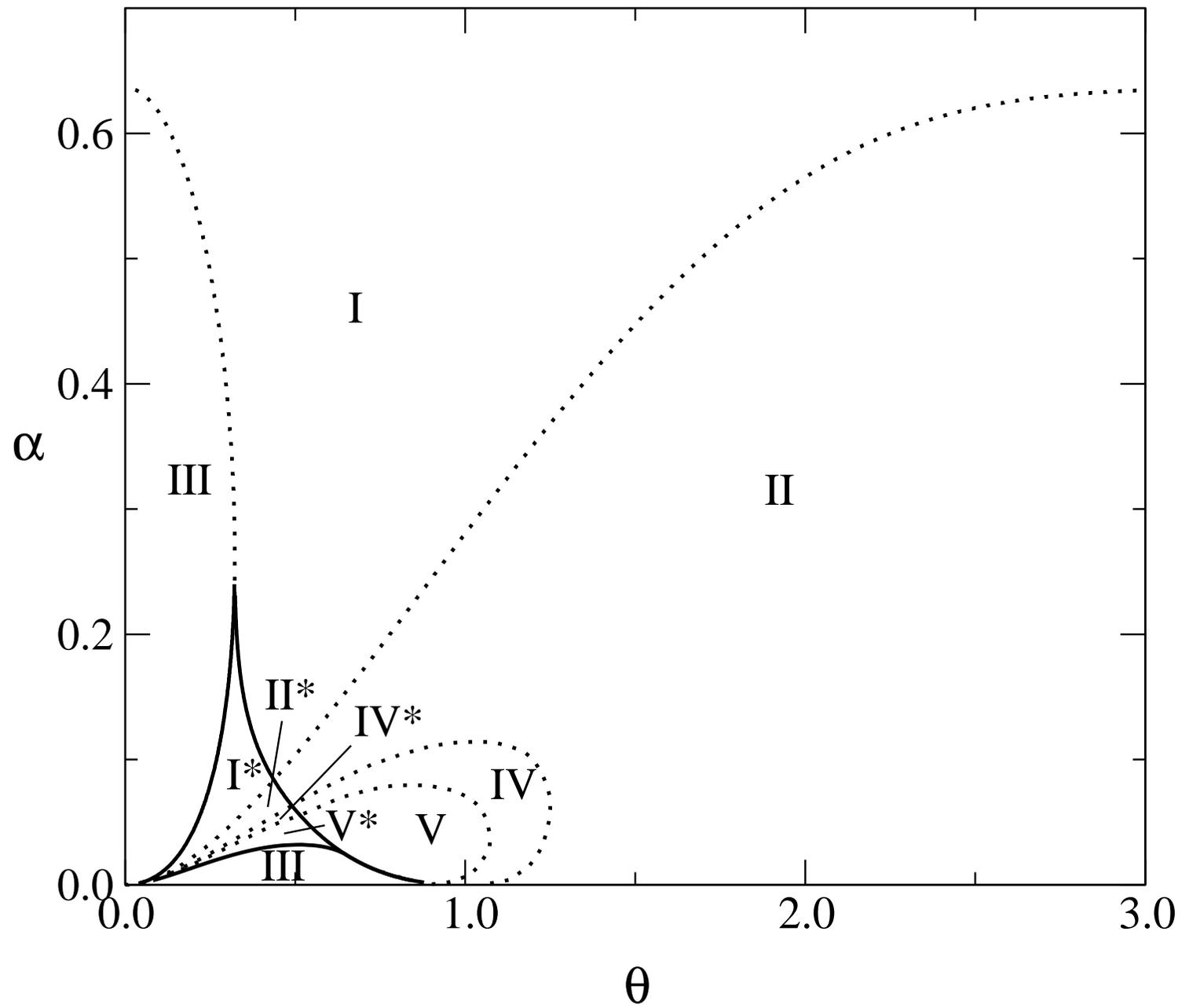
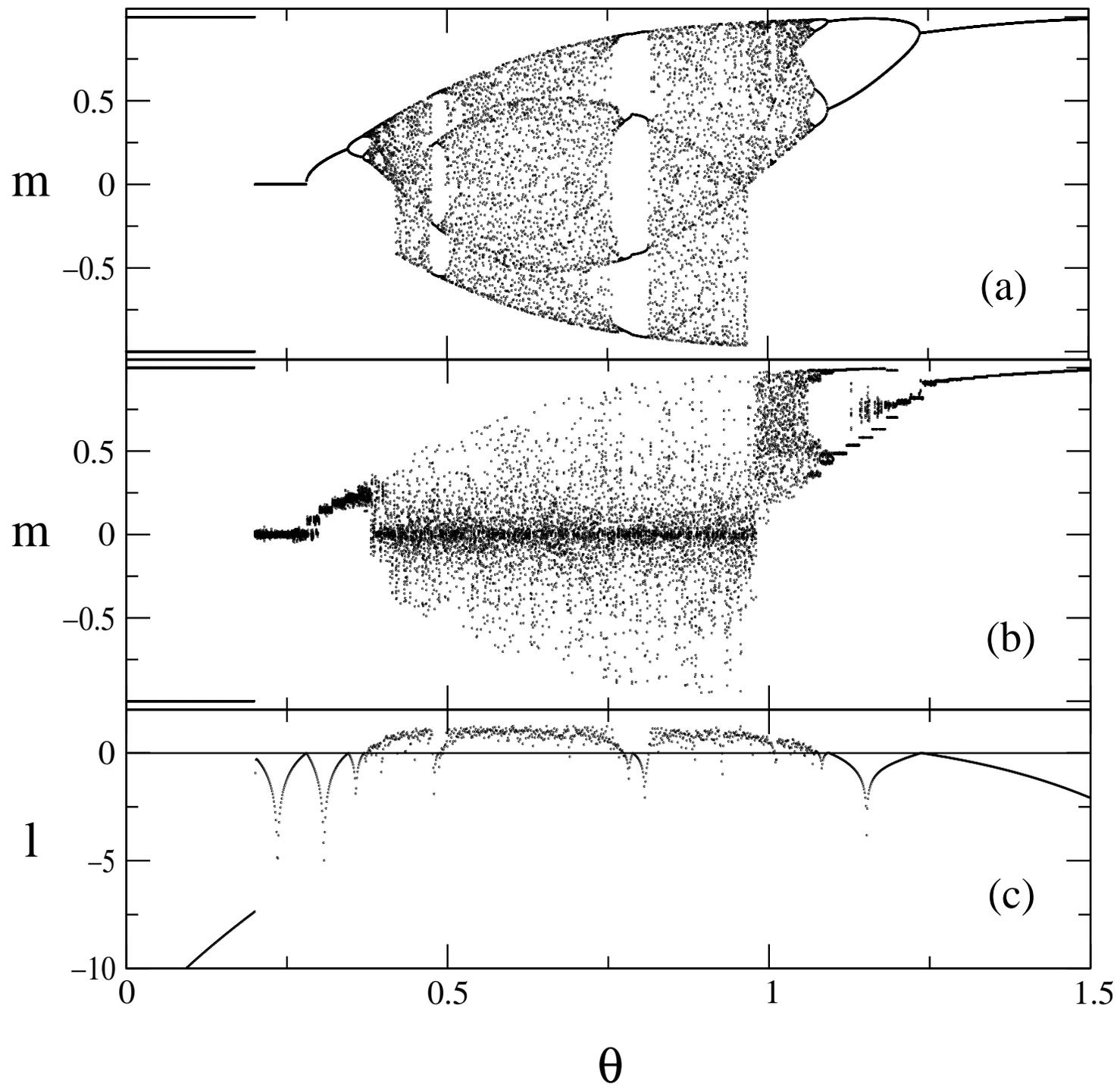
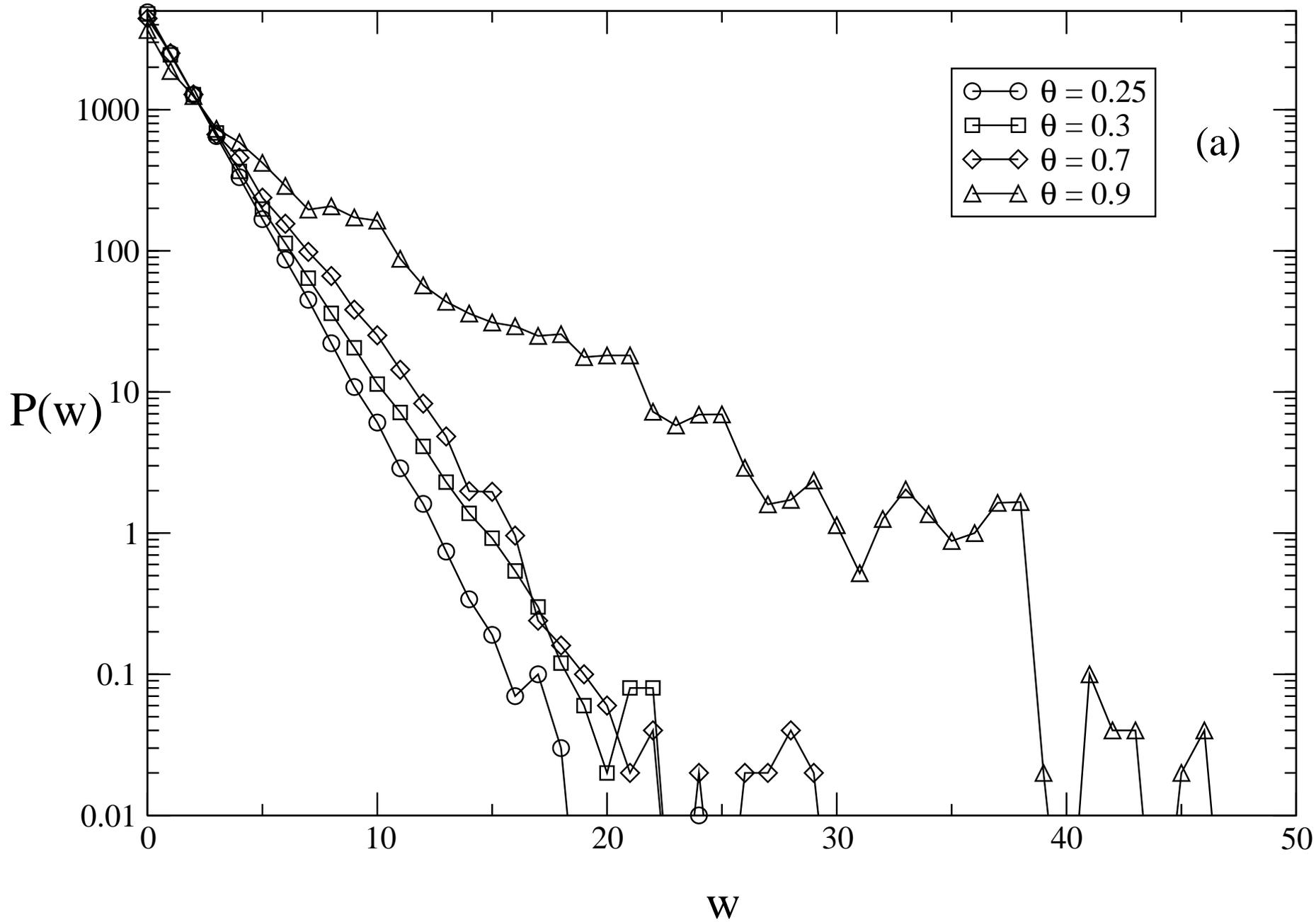
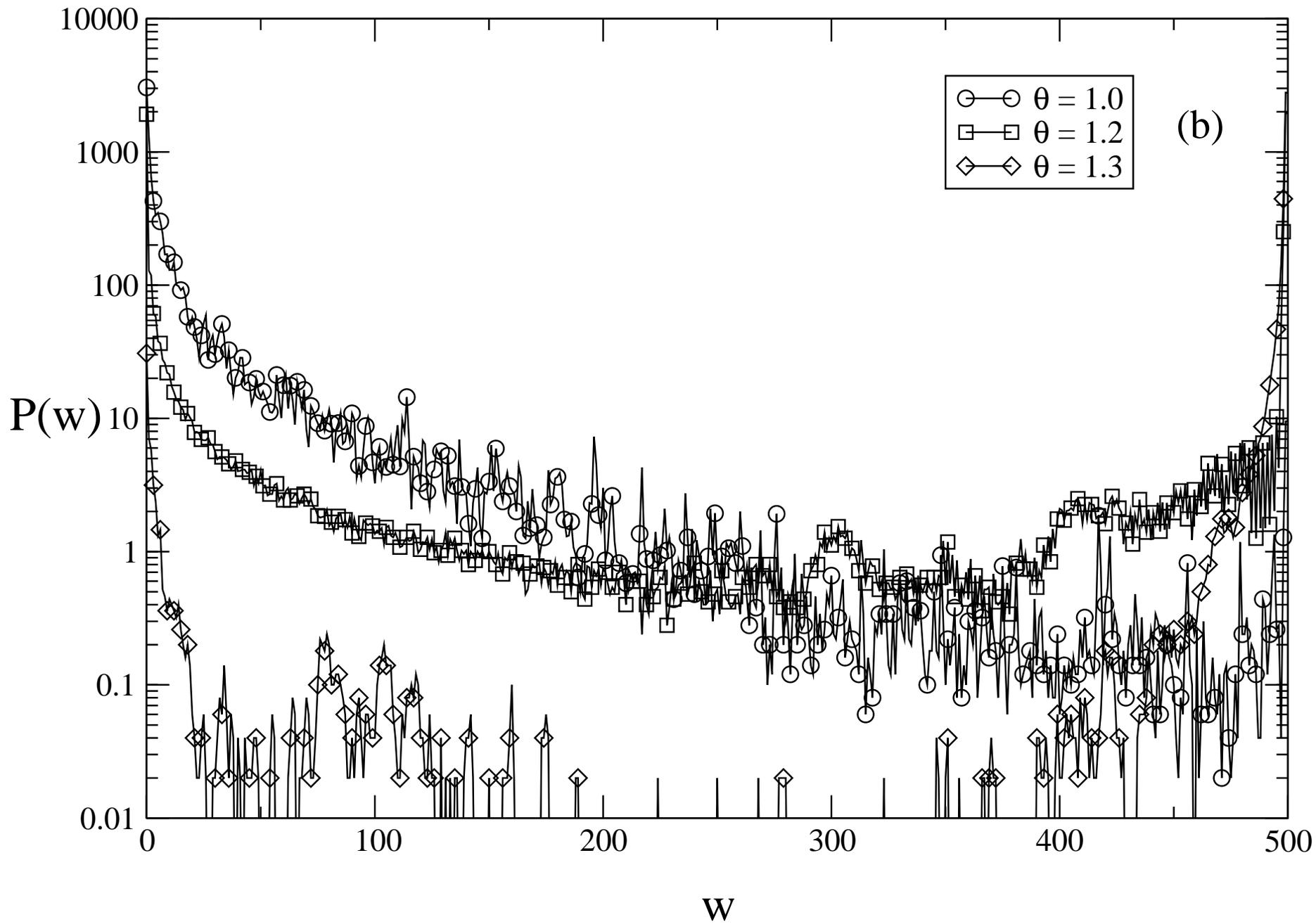
\end{document}